\documentclass[11pt]{article}
\usepackage{psfig}

\begin{document}

\title{Detection of Phase Jumps of Free Core Nutation of the Earth and their
Concurrence with Geomagnetic Jerks}
\author{Toshimichi Shirai$^1$, Toshio Fukushima$^2$, and Zinovy Malkin$^3$ \\[1ex]
\it $^1$Goldman Sachs (Japan) Ltd., Minato-ku, Tokyo, Japan \\
\it $^2$National Astronomical Observatory of Japan, Mitaka, Tokyo, Japan \\
\it $^3$Institute of Applied Astronomy, Russian Academy of Science, St.~Peterburg,
 Russia}
\date{August 2, 2004}
\maketitle

\begin{abstract}
We detected phase jumps of the Free Core Nutation (FCN) of the Earth directly
from the analysis of the Very Long Baseline Interferometer (VLBI) observation
of the Earth rotation for the period 1984-2003 by applying the Weighted
Wavelet Z-Transform (WWZ) method and the Short-time Periodogram with the Gabor
function (SPG) method.  During the period, the FCN had two significant phase
jumps in 1992 and 1998.  These epochs coincide with the reported occurrence of
geomagnetic jerks.
\end{abstract}

\section{Introduction}

The Earth rotation possesses a free nutational mode rotating clockwise
(retrograde) with the period about 430 sidereal days as viewed from the
inertial frame.  This mode, named as the Free Core Nutation (FCN), appears
since the Earth has a rotating and elliptical fluid core (Toomre, 1974).  The
FCN influences the Earth rotation in two different ways; (1)the modification
of forced nutation terms through a indirect response as a non-rigid body, and
(2)the appearance of free oscillation modes as direct effects.  Investigation
of the FCN is important in order not only to improve theoretical modeling of
the nutations but also to understand nature of the Earth interior.  For
instance, the period of the FCN is closely related to the dynamical
ellipticity of the liquid outer core while the quality factor (Q-value) of the
FCN is related to the viscosity of the outer liquid core and the core-mantle
coupling (Sasao, 1980; Getino and Ferrandiz, 2000).

The time variation of the FCN amplitude has been well studied
from the direct effect in the VLBI data (e.g. Shirai and Fukushima, 2001b;
Dehant et al., 2003).  A recent research also suggests that the atmosphere
could have sufficient power to excite the FCN amplitude observed (Dehant et
al., 2003) and huge earthquakes could cause impulsive changes of the FCN
amplitude (Shirai and Fukushima 2001b).  In contrast, most of recent
determinations of the FCN period are based on the estimation of the indirect
effect using VLBI nutation of the Earth rotation (e.g. Shirai and Fukushima,
2001a) or monitoring of the tidal variation of the gravity (e.g. Sato et al.,
1994).  Usually these works assumed the FCN period is as constant value.
Therefore the time variation of the FCN period is unavailable.  Exceptions are
Roosbeek et al. (1999) and Hinderer et al. (2000), who analyzed the time
variation of the FCN period from the indirect effect in the VLBI data.  They
concluded that the FCN period has been stable with a precision of three
sidereal days.  One disadvantage of the determinations based on the indirect
effects is that the time variation of the FCN phase is ignored.  In general,
the analysis of the time variation of the FCN frequency consists of two
factors; that in period and that in phase.  However the time variation of the
FCN phase is not studied yet.  By directly applying the Wavelet Weighted
Z-Transform (WWZ) method and the Short-time Periodogram with Gabor function
(SPG) method to the VLBI nutation data, we revealed the time variation of the
FCN phase and suggest its geophysical cause.

\section{Method of Analysis}

The methods we deployed in the analysis are the WWZ method and the SPG method.
Each method has its own advantages and disadvantages.  The WWZ method was
initially developed for the period analysis of variable stars by (Foster,
1996), where time series of data are unevenly sampled.  One of its merits is
that the WWZ traces the time variation of both the amplitude and period at the
same time.  On the other hands, this method can not treat complex-valued data,
or vector data in general, and does not take care of the data with different
weights.  The WWZ method does not cover all the requirements of FCN analysis,
where the data are usually expressed in a complex value (Shirai and Fukushima,
2001a) as \begin{equation} \zeta_{FCN} = \Delta \psi_{FCN} \sin\epsilon_0 +
i\Delta \epsilon_{FCN}.  \end{equation} Here $\Delta \psi_{FCN}$ is the FCN in
longitude, $\Delta \epsilon_{FCN}$ is that in obliquity, and $\epsilon_0$ is
the obliquity of ecliptic at J2000.0.  Note $\Delta \psi_{FCN}$ and $\Delta
\epsilon_{FCN}$ are two projections components of axis motion of the Earth
rotation in rectangular coordinates viewed from the space.  Also the VLBI data
set contain the information on the weights of observed values of
$\zeta_{FCN}$.  Then its consideration is important for the data before 1990,
when the observation was not so precise.

The SPG method was developed by ourselves to handle unevenly sampled
and complex-valued (i.e., in two-dimension) time series with the weights.
Let us consider some sinusoidal signals embedded in unevenly sampled and
complex observation time series as
\begin{equation}
z_j = x_j+iy_j\end{equation}
with varied observation formal errors, respectively $\delta x_{j}$ and $\delta y_{j}$.
Here the subscript $j$ on values corresponds to those at $t_j$.
Firstly, we assume that the data are associated with the standard weights
\begin{equation}
w_j ={{1}\over{\left(\delta x_{j}\right)^{2}+\left(\delta y_{j}\right)^{2}}}.
\end{equation}
In addition to that, we adopted the Gabor function based on the Gaussian function
as an optimized window function for the time-frequency analysis (Gabor, 1946).
\begin{equation}
G^{\alpha}_{b}\equiv g_{\alpha}(t-b)
\end{equation}
where
\begin{equation}
g_{\alpha}(\tau)={{1}\over{2\sqrt{\pi\alpha}}}e^{-{{\tau^2}\over{4\alpha}}}.
\end{equation}
Here $\alpha$ is a trade-off parameter between time-resolution and frequency-resolution and $b$ is a center epoch for the time shift.
Namely the larger $\alpha$ corresponds to the better frequency resolution.
The time-frequency analysis would be achieved by shifting $b$
discretely as $b_{k} = b_{0}+\Delta_{b}k$ where $\Delta_{b}=b_{1}-b_{0}$.
To obtain periodgram at the specific epoch $b_{k}$, a least square fitting of
sinusoidal curve to the windowed data $wG^{\alpha}_{b_{k}}z$.
Here $\omega_{l}$ is discretely divided frequency
as $\omega_{l} = \omega_{0}+\Delta_{\omega}l$
where $\Delta_{\omega} = \omega_{1} - \omega_{0}$.
Then periodgram $P_{b}^{\alpha}(\omega_{l})$ at specific epoch $b_{k}$
is evaluated as
\begin{equation}
P_{b_{k}}^{\alpha}(\omega_{l}) = {{\rho(\omega_{l})}\over{\sigma(\omega_{l})}},
\end{equation}
where
\begin{equation}
\rho(\omega_{l}) = \sum_{j} G^{\alpha}_{b_{k}}w_{j}z_{j}e^{-i\omega_{l} t_{j}},
\sigma(\omega_{l}) = \sum_{j} w_{j}.
\end{equation}
To trace period changes during observation period, we just pick up the
frequency of the maximum amplitudes from the periodgram at each epochs.
One disadvantage of this method is that the accurate
time variation of the amplitude is unavailable.
Note that the estimated frequency variation would include
not only the time variation of the FCN period but also that of its phase.

We performed a simple simulation with artificial test data to
compare effectiveness of these methods.
We created a test data set of unevenly sampled chirp signals
whose frequency is slowly changing as a linear function of time
($f = f_0 + f_{1}t$).
The timings of sampling were
set the same as those of the actual VLBI data for
the FCN analysis. Of course, we added Gaussian noise of small amplitude.
Figure~1 shows that both the WWZ and the SPG methods
precisely traced the chirp signal.
We confirmed existence of the so-called edge effects, which has already been reported
(Foster, 1996; Malkin and Terentev, 2003),
is observed around the first and last 2 to 3 years.
In the viewpoint of the RMS after fitting, we conclude
that the SPG method is superior to the WWZ method.
In fact, the RMS for the SPG method is 3.4 day,
which is smaller than 4.5 day, that for the WWZ method.
In any sense, the RMS for both the methods are small enough to trace the
time variation of the FCN frequency of the Earth.

\begin{figure}[ht]
\centerline{\psfig{figure=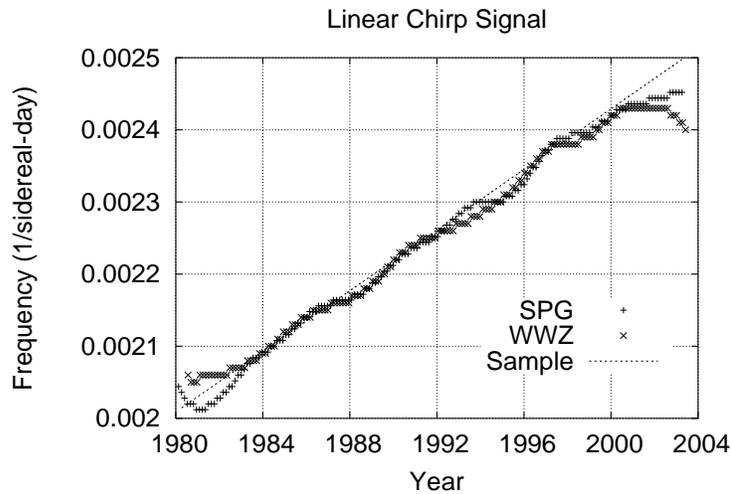,width=0.6\textwidth,clip=}}
\caption{Both the WWZ method and the SPG method precisely traced the chirp signal.
The RMS for the SPG method and the WWZ method are 3.4 day and 4.5 day, respectively.}
\end{figure}

\section{Analysis of VLBI data}

We used the VLBI nutation data complied by the U.S. Naval Observatory for the
period MJD 44089.994 to 52779.206.  Figure~2 shows residuals after subtraction
of a standard model of the forced nutation, the IAU2000A (Mathews et al.,
2002).  Figure~2 clearly illustrates the existence of the FCN (Vondrak, 2003).
Note that nutation data before 1984 will not be used for our later analysis
since they are too noisy and have few data points.  In the case of the WWZ
method, we applied it separately to $\Delta \psi$ and $\Delta\epsilon$ without
weights.  13:27 2004/07/26This is because the WWZ method can not handle with
the vector data with the variable weights as explained.  Figure~3 shows the
large time variation of the FCN frequency.  The result estimated by the SPG
method is slightly different from that by the WWZ method ,however, they are
practically the same, namely with a period difference less than 10 days.  We
suppose that the result estimated by the SPG method is more accurate since the
SPG method considers weights while the WWZ does not.

\begin{figure}[ht]
\centerline{\psfig{figure=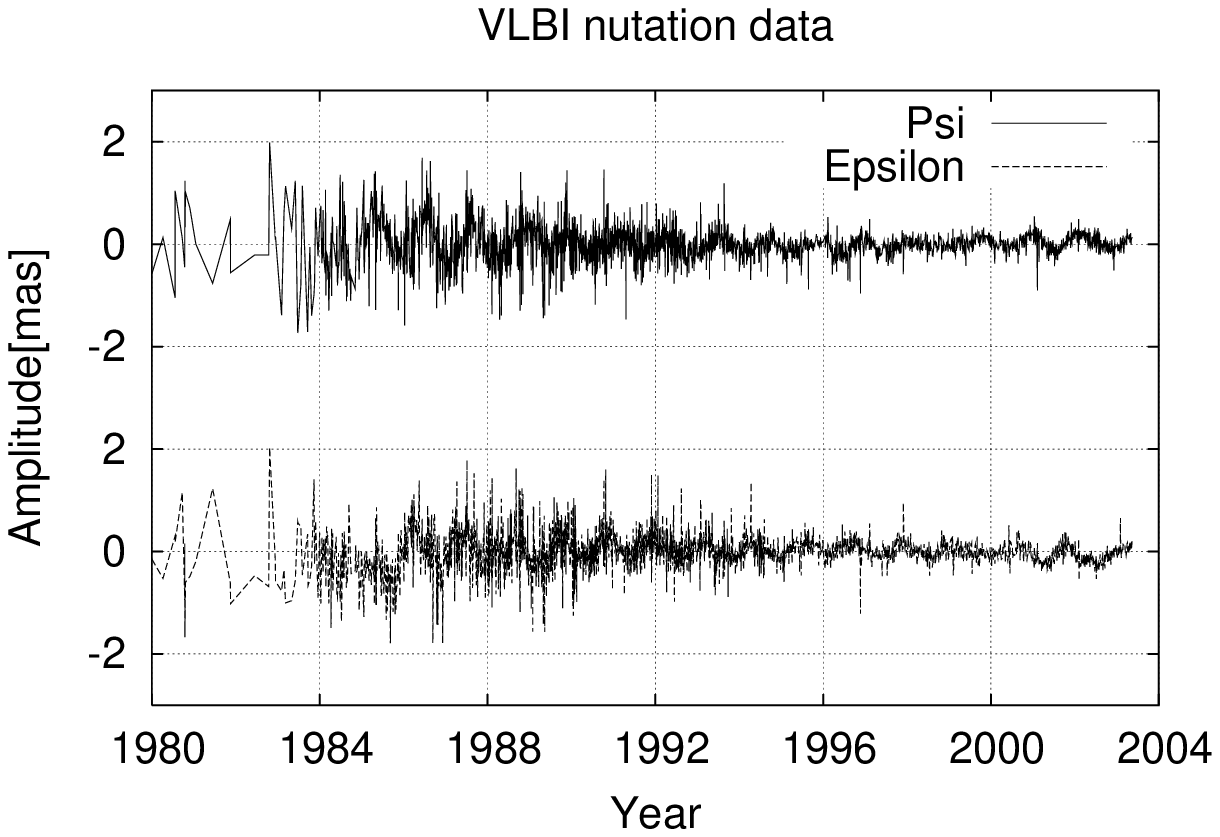,width=0.6\textwidth,clip=}}
\caption{Residuals after subtraction
of a standard model of the forced nutation, the IAU2000A.
Note that nutation data before 1984 is too noisy and has few data point.}
\centerline{\psfig{figure=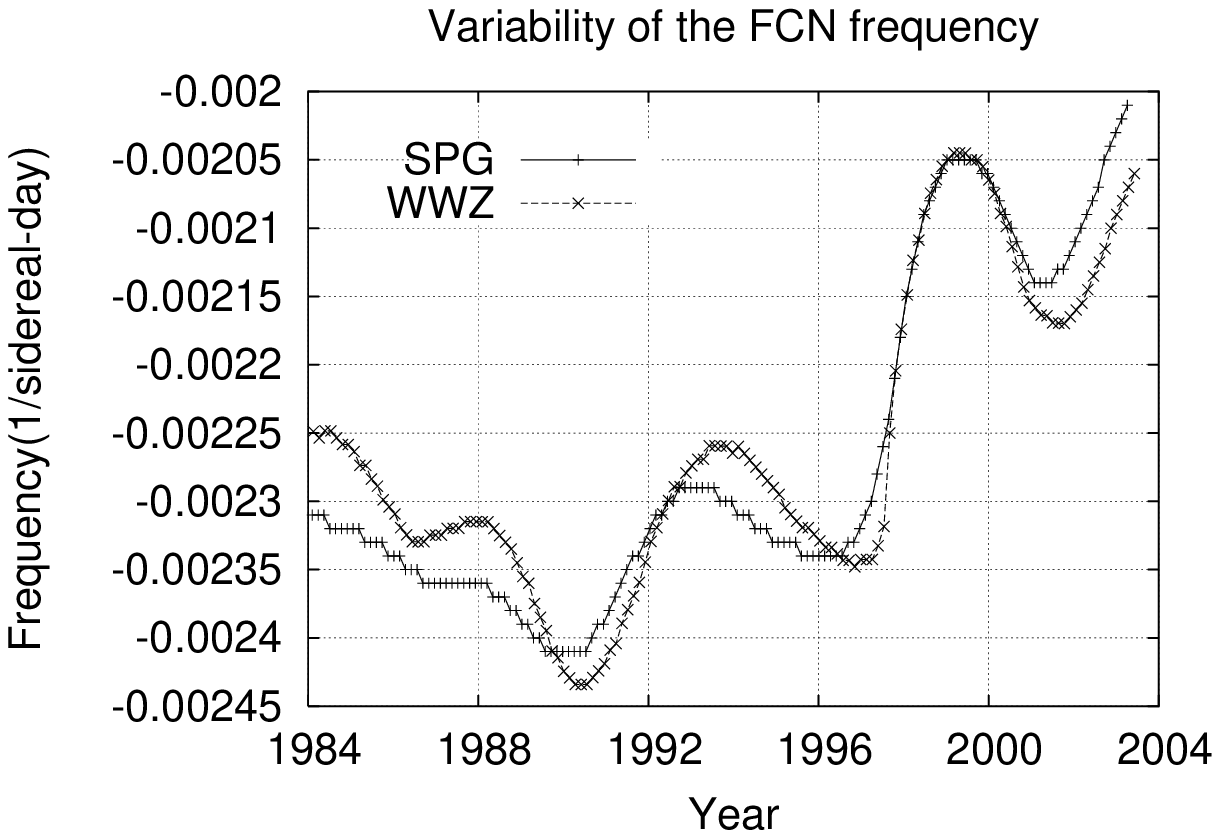,width=0.6\textwidth,clip=}}
\caption{Time variation of the FCN frequency.
The result obtained by the WWZ method is mean
of $\Delta\psi $ and $\Delta\epsilon$
since we applied the WWZ method to them separately.}
\end{figure}

The first thing we need to consider is whether this result is an artifact or
not.  The results of the numerical tests strongly suggests that this large
time variation is not an artifact since the RMS of residuals measured in
period is of the order of several days.  The consistency between the results
estimated by two different approaches, namely the SPG method based on the
periodogram and the WWZ method based on the wavelet transform, also supports
that our result is not an artifact.  Secondly we need to consider possibility
of contaminations caused by excitations of the FCN since neither the WWZ
method nor the SPG method take care of those effects explicitly.  One good
candidate of the FCN excitation mechanism is atmosphere.  It is hard to
estimate the atmospheric effects on our result since the atmospheric model in
diurnal time scale is not well known.  A recent research suggests that the
period of maximum atmospheric excitation varies with time, and it was closest
to the FCN period around 1987 (Dehant et al. 2003).  If the atmospheric
excitation is mainly responsible for these period and/or phase variation, the
maximum time variation should have happened around 1987.  However Figure~2 and
3 show no signs of such effects.  Therefore we conclude that an excitation
effect on our result is limited.

Now let us discuss which factor, namely the period or the phase, cause this
large time variation in the FCN frequency.  As we mentioned before, the FCN
period is closely related to the dynamical ellipticity of the outer liquid
core, which is supposed to be constant at this time scale.  Additionally if
FCN period significantly changes during the observation, the magnitude of the
forced nutation itself must be also affected through the resonant effect.
However the VLBI data show no signs of such effects.  Therefore it is quite
natural to treat the FCN period as a constant and assume that the large time
variation of the FCN frequency mainly comes from that of the FCN phase.

\section{Discussion}

The FCN phase also does not change without forcing changes.  Let us discuss
the geophysical cause of this time variation of the FCN phase.  We note that
ten phase jumps were observed in the Chandler Wobble (CW) spanning the years
1890-1997 whose durations are between 1 to several years (Gibert et al. 1998).
The CW is another free oscillation of polar motion of the Earth and their
period is close to 435 mean solar days as viewed from the terrestrial frame
(Lambeck, 1980).  In detecting the phase jumps, the authors assumed a simple
model for the phase variation as \begin{equation} \Delta \phi = \sum_{k=1}^N
a_k \psi \left(\frac{t-t_k}{2^{1/2}\delta a_k} \right) \end{equation} where
\begin{equation} \psi(\tau)= \frac{1}{2} + \frac{1}{\pi} \int_{0}^{\tau}
e^{-t^2} dt.  \end{equation} Here is $N$ the number of phase jumps, $t_k$,
$\delta_k$, and $a_k$ are the mean time, the characteristic duration, and the
amplitude of the $k$-th phase jump, respectively.  The largest phase jump
happened in 1925.  It amounted to $152^\circ$ in angle.  The broadest duration
of the phase jump happened in 1953 and amounted to 15 years.  The above
authors also found that the occurrence of the phase jumps follow those of
geomagnetic jerks with a delay not exceeding three years.  The geomagnetic
jerk is a rapid change in rate of the secular variation curve of geomagnetic
fields.  On the other hands, the theoretical model suggests that the
instability of a layer at the top core and its downward propagation induce a
step in the core-mantle torque strong enough to explain the phase jumps in the
CW (Bellanger et al., 2001).

To reveal the time variation of the FCN phase, we calculated it from the time
variation of the FCN frequency by fixing the FCN period as a constant
determined from the indirect effect, namely 430 sidereal days (Shirai and
Fukushima, 2001a).  Namely we write the deviation of the phase from a linear
function of time as \begin{equation}
\Delta\phi(t)=\phi(t)-2\pi\nu_{FCN}(t-t_0) \end{equation} where
\begin{equation} \phi(t)=\int_{t_0}^{t}2\pi\nu(t)dt.  \end{equation}
Here
$\nu(t)$ is the observed time variation of the FCN frequency shown in Figure~3,
$\nu_{FCN}$ is the constant FCN frequency, and $\Delta\phi(t)$ is the time
variation of the FCN phase.  This transform means that we assume that the
argument of the circular functions in the functional expression of the
original signal is cast as a sum of a linear and a time variable phase instead
of assuming that the original signal has the function form similar to a
harmonic oscillation but with the angular frequency not constant but function
of time.  Figure~4 shows $\Delta \phi (t)$ for the last 20 years.  The figure
indicates that there are two sudden trend changes of the FCN phase in 1992 and
1998.  Before and after those sudden changes in trend, the time variation of
the FCN phase is approximated by a the linear function or ERF function.  This
characteristic is explained by the same mechanism on the phase jumps in the CW
described as Eq(8) though the numbers of jumps are so different, two versus ten.

\begin{figure}[ht]
\centerline{\psfig{figure=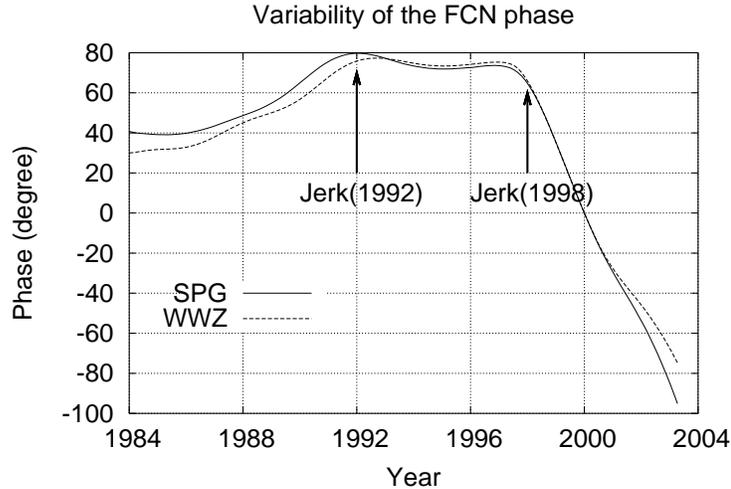,width=0.6\textwidth,clip=}}
\caption{Time variation of the FCN phase.
The FCN phase has two sudden trend changes in 1992 and 1998.
Two arrows plotted present epochs of geomagnetic jerks.}
\end{figure}

One explanation is that at least one phase jump at a long duration
occurred before the observation period which is responsible for the phase
variation before 1992.  Unfortunately an accurate estimation of the parameters
of each phase jumps is not feasible due to the limited observation period.
This is mainly because the impacts of the phase jumps happened before 1992 and
in 1998 remain beyond the observation period.  Without those accurate
estimations, it is difficult to perform accurate estimations of the parameters
of the phase jump happened around 1992.  On the other hands, the geomagnetic
jerks occurred twice during the observation period, namely around 1992 and
1998 or 1999 (Huy et al., 1998; Huy et al., 2000; Mandea et al., 2000).  These
epochs coincide the dates of the phase jump of the FCN as well as the CW.
Regarding a candidate of the phase jumps before the observation period, the
geomagnetic jerk occurred in 1979 (Huy et al., 1998) while the recent research
suggests the other might have occurred in 1983 too (Wardinski et al, 2003).

\section{Conclusion}

We detected the phase jumps of the FCN of the Earth directly from the analysis
of the VLBI observation for the period 1984-2003.  During the period, the FCN
had two significant phase jumps in 1992 and 1998.  These epochs coincide with
the occurrence of the geomagnetic jerks.  We have only two geomagnetic jerks
during the limited observation period and its theoretical model has not been
unavailable yet.  On the other hand, around ten phase jumps in the Chandler
wobble are reported to have occurred in the period 1870-1997.  Those epochs
are also close to those of the geomagnetic jerks (Gibert et al., 1998;
Bellanger et al., 2002).  A theoretical model on such phase jumps is already
provided by (Bellanger et al., 2001).  We regard that it is important to
analyze the VLBI nutation data of a longer observation period and construct a
theoretical model as well as the CW.  The geomagnetic jerk is the rapid change
in rate of the secular variation curve of one of geomagnetic field.  Since its
origin is supposed to be inside of the Earth, it is surely an important issue
for the study of the dynamic of the Earth interior, especially the study of
the conductivity properties of the mantle.  Therefore this phenomena could be
a new diagnostic tool for the investigation of the Earth's interior through
the VLBI data.  Lastly we think it worths to mention that a trend towards the
Earth's dynamic oblateness $J_{2}$ also changed suddenly after 1998 (or 1999)
(Cox and Chao 2002).  Although its geophysical cause(s) are uncertain, they
mention the geomagnetic jerk in 1998 as the potential geophysical cause.
Windows application of the SPG method is available from the author TS.

\section*{Acknowledgments}
We gratefully thank for Ferrandiz, J. M. and Oliver, de Viron for valuable
comments on this paper.

\end{document}